# *Ab initio* Calculation of Dipole Moments and Transition Dipole Moments of $HCl^+$ and $HBr^+$ Molecular Ions

**Valerij S. Gurin[a], Mikhael V. Korolkov[b]**
[a]*Research Institute for Physical Chemical Problems,*
*Belarusian State University, Minsk, Belarus*
[b]*Institute of Physics, NANB, Minsk, Belarus*

**Abstract**

Electronic structure of $HCl^+$ and $HBr^+$ molecular ions is calculated using the symmetry-adapted-cluster configuration interaction (SAC-CI) method. In this paper, we analyse dipole moments (DM) for a series of low-lying six $^2\Pi$-states and transition dipole moments (TDM for transitions from the ground state $X^2\Pi$ to the excited $^2\Pi$-series. Behavior of DMs with change of interatomic distances is different for these ions for the excited $^2\Pi$-states in correspondence with different dissociation paths. TDMs reveal the pronounced maxima at the beginning steps of dissociation.

**1. Introduction**

In the recent paper [1] we have described principal results of *ab initio* calculations for $HCl^+$ and $HBr^+$ molecular ions. We have used the SAC-CI method that allows treatment of the ground state and a series of excited electronic states with high accuracy. In the work [2] we have presented the SAC-CI calculations of $HCl^+$ with detailed comparison with previous research by Dalgarno et al [3] and experimental data available. In general, there is rather good consistence (both experiment and other calculations) for the first three states of $^2\Pi$-symmetry ($X^2\Pi$ is the ground state for both $HCl^+$ and $HBr^+$ ions) and the lowest states of other symmetries: $^2\Sigma^+$, $^2\Sigma^-$ and $^2\Delta$ series. Part of these states are bound ones, and the comparison with calculations appears to be quite well [2,3]. The potential curves for three $^2\Pi$-states for $HCl^+$ and $HBr^+$ are very similar, however, new features begin from the 4[th] states those enter the next energy interval (about 10 eV higher). These higher states demonstrate also bound behavior at the distances less than the equilibrium distance of the corresponding ground state. It should be expected that non-trivial behavior will be for other properties of the ions by monitoring their R-dependencies for a series of excited state. In the present paper, we concern dipole moment (DM) and transition dipole moment (TDM) for series of $^2\Pi$-states. Electronic transitions can occur from the ground state, $X^2\Pi$, to the higher ones. Values of DM results in various interaction of the ions with electromagnetic field, and different modes of dissociation ($H^+ + X^0$ and $H^0 + X^+$) provide the asymptotics to zero or non-zero constant due to unbound atoms.

Calculations of electronic structure of neutral molecules and molecular ions at this level need taking into account many excited states, together with the data on the other molecular properties, DM, TDM and vibrational spectra. In spite of great attention throughout years [4-10], there is a deficiency in knowledge of excited electronic states of neutral and ionic hydrogen halides, $HCl^+$ and $HBr^+$. Meanwhile, a number of excited states (at least 4-5 states) need to describe adequately photochemistry of the molecules under powerful laser excitation. Within the framework of this study we consider two molecular ions from this series, $HCl^+$ and $HBr^+$. These molecular ions possess the doublet ground states, and there are no stable closed shell states. Open electronic shells dictate requirements to use high-level *ab initio* calculation methods with configuration interaction (CI). Up to date, there is no sufficient knowledge on a series of the states of $HCl^+$ and $HBr^+$ in the range of energies up to 5-6 eV above the ground state (i.e. the range of optical excitations). $HCl^+$ has been studied much more than $HBr^+$, however, the excited states above the 3$^{rd}$ one are worth to be recalculated and discussed more in detail. We have presented the data for several states of $HCl^+$ [1,2] using the advanced *ab initio* calculation scheme, the method of symmetry-adapted-clusters configuration interaction (SAC-CI) [10,11]. The SAC-CI method has been developed for calculations of ground and excited states of molecular systems admitting a tuning of the calculation accuracy in correspondence with computing performance and system complicacy. This calculation scheme takes an optimum cluster expansion for electronic excitations with selection of expansion terms based on symmetry using optimized selection of the reference states to account CI of one- and multi-electron excitations allowing to attain good accuracy for joint analysis of many excited states. The results are at the level of another popular multiconfiguration quantum chemical calculation approaches like CASSCF, CCSD, MRSDCI, etc.

Within the framework of this publication we have restricted ourselves to a series of $^2\Pi$-states without accounting the spin-orbital effects (those are small $HCl^+$, but noticeable for $HBr^+$) and continue our analysis of electronic structure of hydrogen halide ions to a series of electronic transitions between these states. $^2\Pi$-states for these ions are of the first interest as they are responsible for photoexcitation processes. The states of other symmetries are the subject of subsequent works.

## 2. Calculation details
For electronic structure calculations of $HCl^+$ and $HBr^+$ ions, SAC-CI method was used at the general-R level including R-operators up to 3rd order accounting the multiple excitations. The details were described in [1,2]. Briefly, with GAUSSIAN software [12] we use the all-electron basis sets of 6-311G quality (Cl atom) with additional diffuse functions and aug-cc-pVDZ (Br atom) providing good consistence with previous data. In the CI procedure for the external electronic shell was treated as active orbitals, internal ones were frozen. The active space for CI included the valence orbitals beginning from and 1s- for H, 3s- for Cl, 4s- for Br consisting of 4 occupied and 58 virtual orbitals.

The values of DM were calculated by the standard procedure (the conventions, e.g. in [13]) integrated into the software used [12]. In order to define DM within this approach, the weight centre of nuclear charges is taken as the origin of coordinates. For ionic diatomics this choice is of importance for comparison with data of other calculations, and may be transformed by simple relations, e.g. to the origin of coordinate system based on the centre of mass.

## 3. Calculation results

The calculated potential curves have been shown in [1,2] to be rather similar for similar for $HCl^+$ and $HBr^+$ that is understandable since the ions are isoelectronic for outer shells. However, the asymptotic behavior at large separations reveals the difference beginning from the first excited states $2^2\Pi$ and $3^2\Pi$ since the dissociation paths for this couple of states are not same. A charge may be either at halogen atom, $H^0(^2S)+Hal^+(^1D)$, or at hydrogen, $H^+ + Hal^0(^2P)$. Evidently, DM values are to be very different for the two versions of dissociation paths and a behavior of DMs is expected to be nontrivial. This is depicted in in Fig.1-2 for the first 6 states of $^2\Pi$-symmetry. For the lowest $^2\Pi$-states of $HCl^+$ these results are in good consistence with calculations in [3]. The two types of DM asymptotics can be derived easily using the classical electrostatic formulae (however, in the coordinate system of centre of masses): a large value for asymptotic $\sim(35/36)eR$ for the case of dissociation to $H^+$ and Cl, and a small value $\sim(1/36)eR$ for the dissociation to H and $Cl^+$ in correspondence with the atomic masses of elements. For $HBr^+$ this derivation is analogous, and the values for asymptotic behaviour are $\sim(80/81)eR$ and $\sim(1/81)eR$, respectively.

At medium separations the dependencies of DMs reveal extreme behaviour with the noticeable maxima at the distances larger than the equilibrium one, and for the first excited state, $2^2\Pi$, for $HBr^+$ this distance is more than $6a_0$. These observations mean that DMs for such states grow during the starting dissociation steps rather than at the equilibrium bound molecules. This can be of importance for interaction of these molecules with external fields, i.e. a dissociating molecules are more susceptible for external fields.

A series of higher states, $4^2\Pi$-$6^2\Pi$ (Fig.1-2) demonstrate more complicated form of R-dependencies those are not similar for both ions under study. It corresponds to different asymptotics and different dissociation paths for these ions. There are several maxima in R-dependencies for all these states, and again the maxima enter the range of dissociating ions. Also, it is remarkable the inversion of sign DMs in the range of short distances for all excited states and in the range of $7$-$9a_0$ for $6^2\Pi$ state. This inversion corresponds to charge redistribution in the course of different dissociation step and can be explained by a strong bond rebuilding in the ions, and the dissociation is accompanied by complex changes in bonding of the atoms rather than any single-bond breaking.

In order to analyse the calculation data for TDMs obtained for $HCl^+$ and $HBr^+$ we concern here the transitions from the ground state, $X^2\Pi$, to three excited ones of the same symmetry (Fig. 3). There are the pronounced maxima for all of them at different separations that is of interest from the point of view of potential excitation and dissociation paths control. For $HCl^+$ the maximum value of TDM occurs for the lower final state, $2^2\Pi$. It is located at the larger distances than for the transition to $3^2\Pi$ final state. This result can mean the possibility of selective excitation of this ion from the ground $X^2\Pi$ state to different final states. The situation of $HBr^+$ is inverted, as well that occurred for DM behavior above. Both maxima of TDM are placed at the distances essentially more than the equilibrium, $R_0$ for the corresponding ground $X^2\Pi$ states for the ions. Meanwhile, at $R_0$ the values of TDM for all these transitions are much lower. Thus, excitation of the ions is much more probable at the beginning dissociation steps rather than at the equilibrium state. However, at the large distances, $R>12a_0$, TDM for all transitions become practically zero. It should be noted that the corresponding potential curves attain a plateau [1,2] at the less distances, $R>8a_0$, i.e. intensive transitions (in particular, for the case $X^2\Pi->2^2\Pi$ for $HCl^+$ and $X^2\Pi->3^2\Pi$ for $HBr^+$) can appear even for weakly interacting atoms (ions) at considerable late dissociation steps.

TDMs from the ground state to $4^2\Pi$ state also have nontrivial behavior with extreme form of R-dependency. However, the similarity between the values for $HCl^+$ and $HBr^+$ remain in the case of this type of transition that has not been presented earlier to our best knowledge. This dependency is similar for $HCl^+$ and $HBr^+$ cases. It is remarkable that TDM to $4^2\Pi$ state changes the sign at $R\sim5a_0$ in this plot that can be considered as a consequence of wave functions inversion while a physical value of intensity of transitions (squared) remains to be positive.

4. Conclusions
*Ab initio* calculations using the SAC-CI method have been performed for study of a series of six doublet $^2\Pi$ electronic states of $HCl^+$ and $HBr^+$ and R-dependencies of DM and TDM have been analysed. DMs demonstrate various asymptotic behavior for the different dissociation pathways, and the pathways are not the same for corresponding excited states of $HCl^+$ and $HBr^+$. There are featured picture at short distances and pronounced maxima at approximately twice larger distances than the equilibrium one. TDMs for the transitions from the ground state to a series of $^2\Pi$ ones attain considerable large values at the distances more than $R_o$ and vanish at separation of the atoms and the positions of maxima are not same for $HCl^+$ and $HBr^+$ in accordance with the difference in dissociation pathways of the corresponding states.

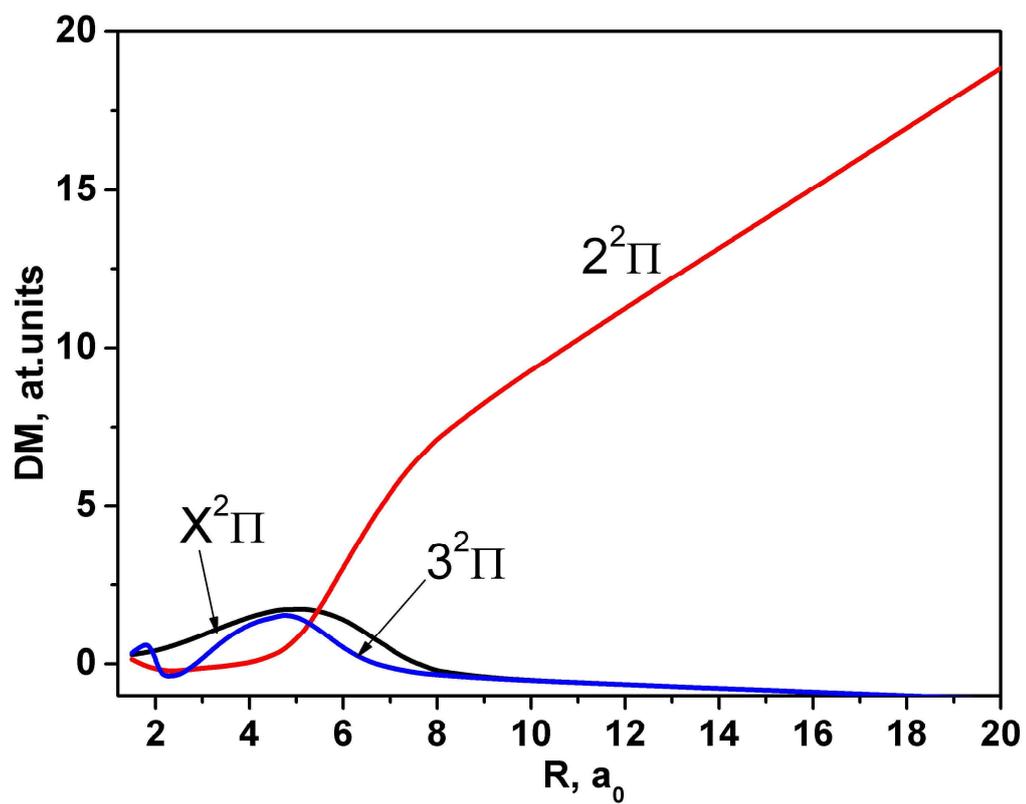

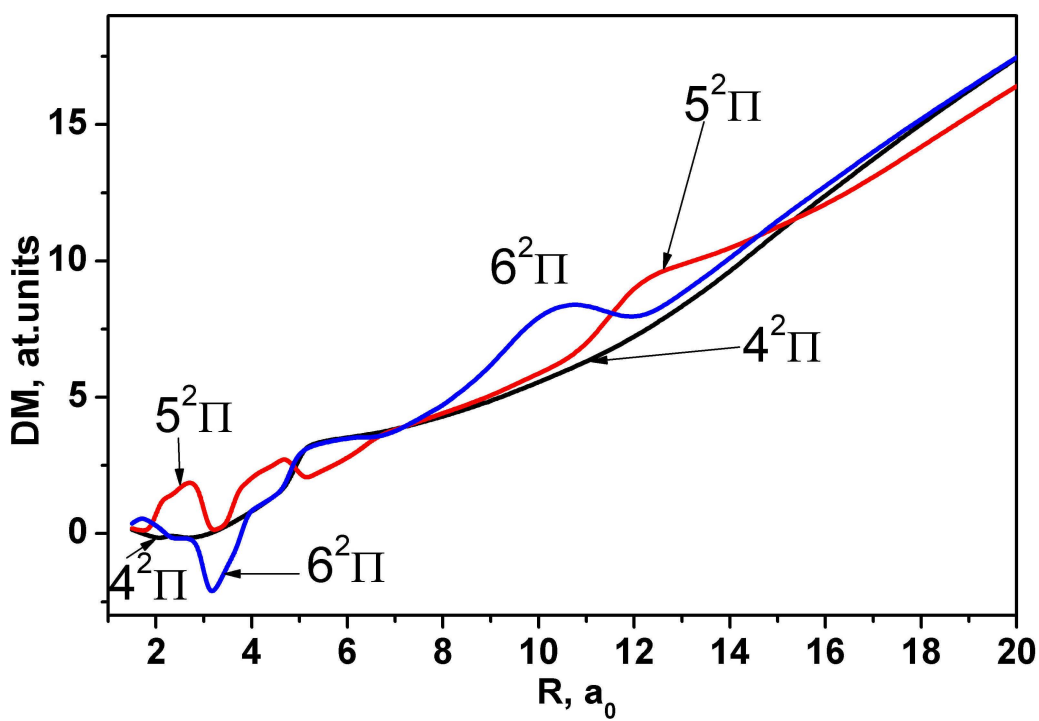

Figure 1. DM functions for a set of six first $^2\Pi$-states of HCl$^+$

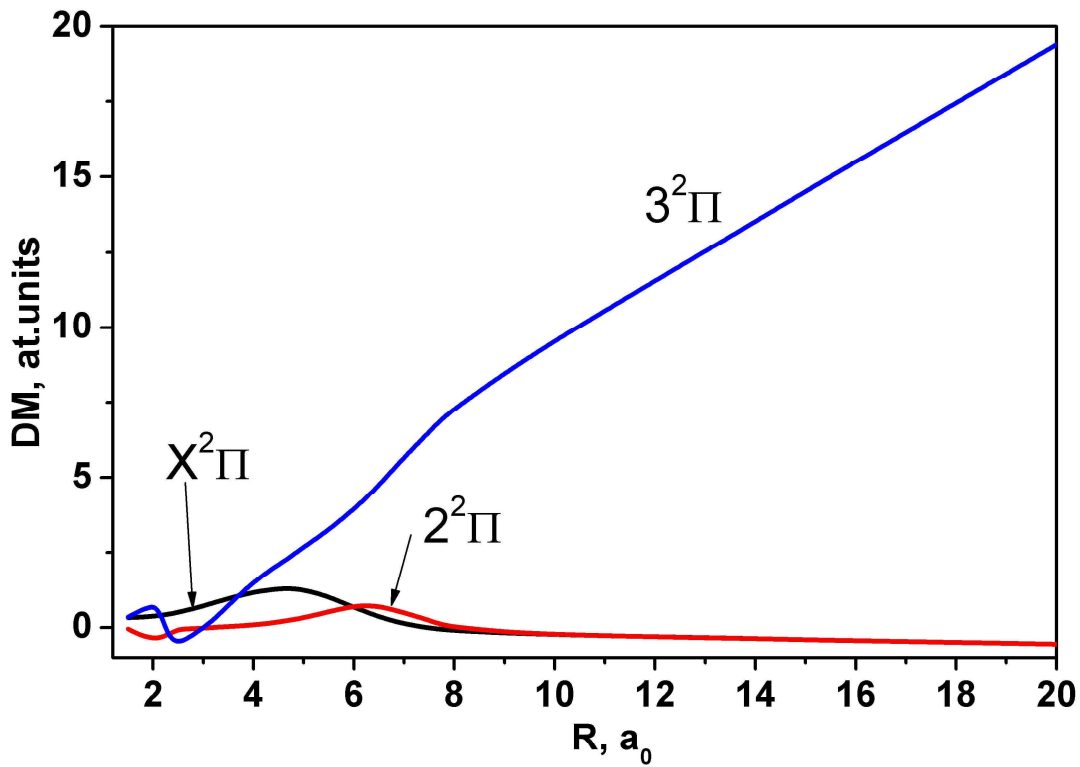

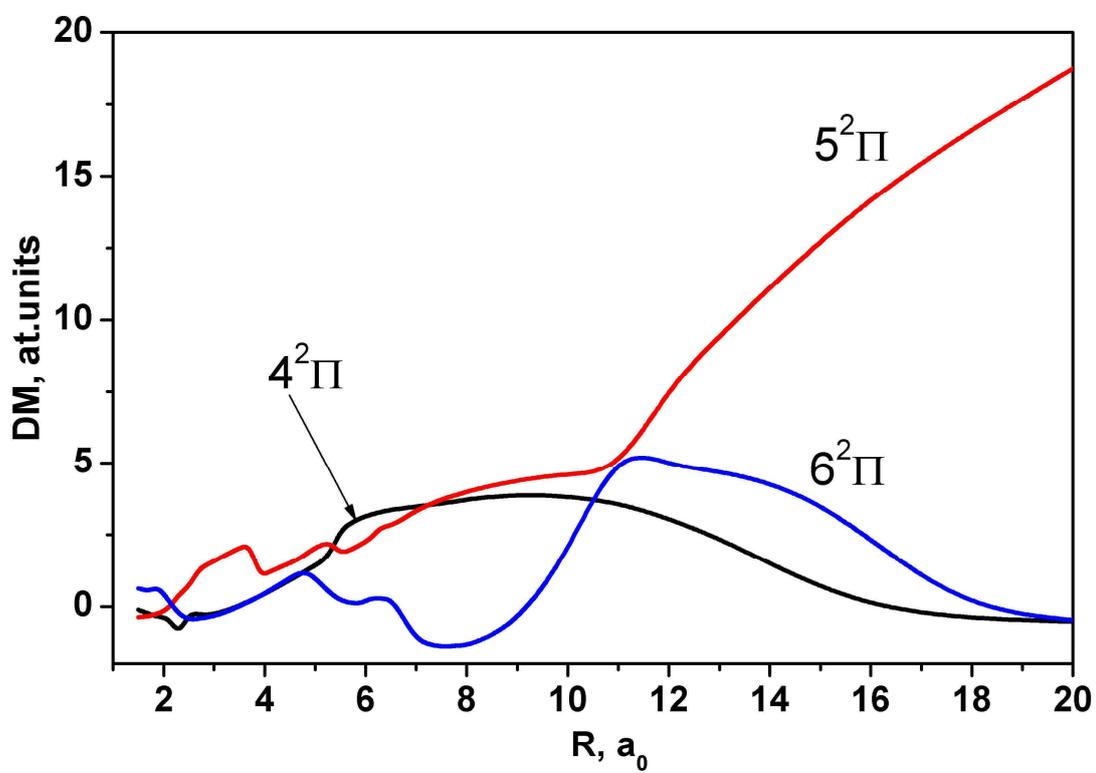

Figure 2. DM functions for a set of six first $^2\Pi$-states of HBr$^+$

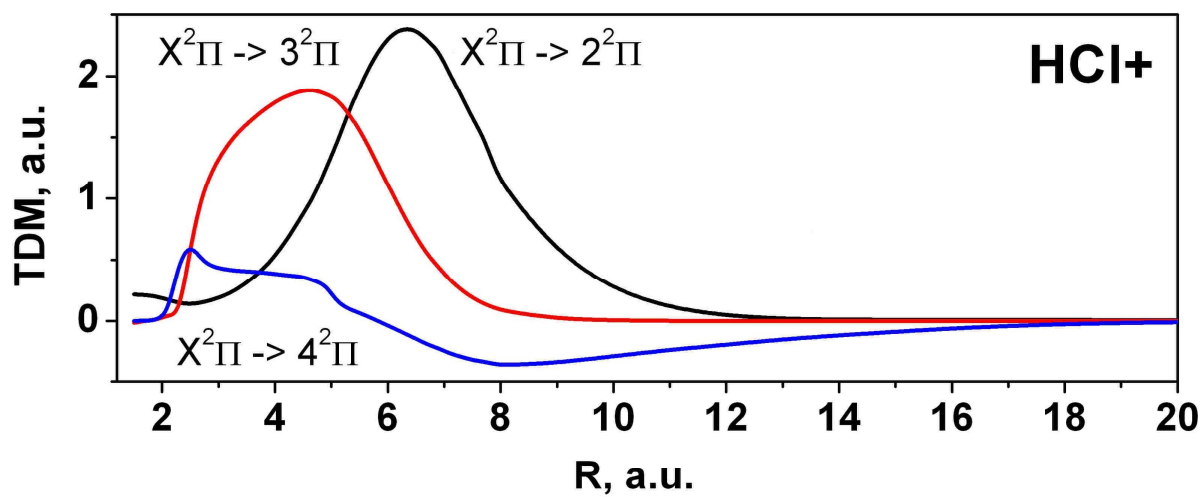

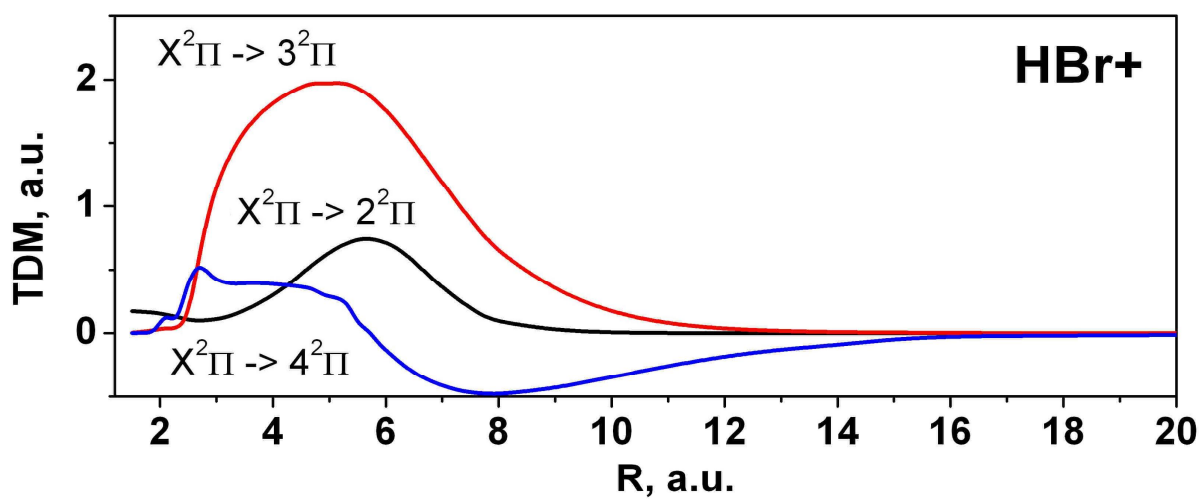

Figure 3. TDM for the transitions from the ground state to the next three $^2\Pi$-states of $HCl^+$ and $HBr^+$